\documentclass[aps,prd,preprint,superscriptaddress,amsmath,amssymb,showpacs]{revtex4-1}
\usepackage{dcolumn}
\usepackage{graphicx}
\usepackage{float}
\usepackage{xcolor}
\usepackage{physics}
\usepackage[colorlinks=true,allcolors=blue]{hyperref}
\begin{document}
	\title{Quark anomalous magnetic moments and neutral pseudoscalar meson dynamics with three-flavor NJL model in magnetized quark matter}
	
	\author{Chang-Yong Yang}
	\affiliation{College of Mathematics and Physics, China Three Gorges University, Yichang 443002, China}
	
	\author{Sheng-Qin Feng}
	\email{Corresponding author: fengsq@ctgu.edu.cn}
	\affiliation{College of Mathematics and Physics, China Three Gorges University, Yichang 443002, China}
	\affiliation{Center for Astronomy and Space Sciences and Institute of Modern Physics, China Three Gorges University, Yichang 443002, China}
	\affiliation{Key Laboratory of Quark and Lepton Physics (MOE) and Institute of Particle Physics,\\
		Central China Normal University, Wuhan 430079, China}
	
	
\begin{abstract}
We investigate the influence of quark anomalous magnetic moments (AMMs) on the mass spectra of neutral pseudoscalar mesons ($\pi$, $K$, $\eta$, $\eta^{'}$) under external magnetic fields, finite temperatures, and quark chemical potentials using the three-flavor Nambu-Jona-Lasinio model. By incorporating AMMs at the quark level, we reveal that AMMs significantly alter the magnetic field dependence of constituent quark masses, inducing first-order phase transitions for light quarks at critical fields, while strange quarks exhibit nonmonotonic mass behavior. The inclusion of AMMs reshapes the QCD phase diagram, suppressing chiral transition temperatures, and the strong magnetic field shifts critical endpoints toward lower $\mu$ and higher $T$ without AMMs. The crossover phase transition without AMMs is replaced by a first-order transition with AMMs under strong fields. Moreover, the inverse magnetic catalysis (IMC) induced by the introduction of AMMs qualitatively aligns with the predictions of lattice QCD (LQCD) for the dependence of phase transition temperature on the magnetic field. For mesons, a larger AMM triggers abrupt mass collapses and enhances flavor spitting at zero $\mu$ and $T$ and accelerates chiral restoration for $K$ and $\eta$ mesons via thresholds tied to strange quark masses in finite $\mu$ and $T$.  These findings underscore AMMs' critical role in reconciling effective model predictions with LQCD results, particularly in reproducing IMC and explaining phase transition dynamics.
\end{abstract}
	
	\maketitle
	
\section{Introduction}\label{sec:Intro}

Strong magnetic fields play a pivotal role in numerous extreme physical environments. For instance, ultrastrong magnetic fields ($\sim10^{23} G$) that may have existed in the early Universe~\cite{Grasso:2000wj,Vachaspati:1991nm}, the extreme magnetic field environments on the surface ($\sim10^{15} G$) and interior ($\sim10^{18} G$) of magnetars ~\cite{Duncan:1992hi,lai1991cold}, and the instantaneous strong magnetic fields ($\sim10^{18} G$) generated in relativistic heavy-ion collision experiments such as RHIC and LHC ~\cite{Tuchin:2013ie,Skokov:2009qp,Guo:2019joy,She:2017icp,Zhong:2014sua,Deng:2012pc,Chen:2019qoe}. In these scenarios, the nontrivial effects of magnetic fields on quantum chromodynamics (QCD) matter have garnered widespread attention, particularly their modulation of the phase structure and dynamical behavior of strongly interacting matter. Specifically, the presence of strong magnetic fields has led to the emergence of numerous new and observable physical effects in quark matter, which contribute to a better understanding of QCD, the fundamental theory of strong interactions, considered one of the most promising theories. For instance, some of these effects involve magnetic catalysis (MC) ~\cite{Shovkovy:2012zn} and inverse magnetic catalysis (IMC) ~\cite{Preis:2012fh} related to the dynamical chiral symmetry breaking mentioned in this article, while others involve related phenomena, such as the chiral magnetic effect and chiral vorticity effect~\cite{Fukushima:2012vr,Kharzeev:2013ffa,Bali:2011qj}.

From a theoretical perspective, magnetic fields significantly affect the properties of QCD vacuum and matter. For instance, the MC effect indicates that magnetic fields can enhance chiral symmetry breaking through dimensional reduction mechanisms, leading to an increase in quark condensation and a rise in phase transition temperature ~\cite{DElia:2010abb,DElia:2011koc,Gusynin:1994re,Gusynin:1994va,Gusynin:1994xp}.

However, lattice QCD (LQCD) simulations at finite temperatures reveal an opposite trend: near the phase transition region, chiral condensation exhibits nonmonotonic behavior with increasing magnetic fields, and the phase transition temperature decreases as the magnetic field increases ~\cite{Bali:2011qj,Ding:2020inp,Bali:2012zg,Ilgenfritz:2013ara}. This phenomenon is known as IMC ~\cite{Bornyakov:2013eya,Bali:2014kia,tomiya2019phase,Andersen:2014xxa}.  The microscopic mechanism underlying this anomalous behavior remains incompletely elucidated, posing a significant challenge for current theoretical research.

To explain the IMC effect, researchers have proposed various phenomenological schemes, including introducing a magnetic-field-dependent coupling constant in the Nambu-Jona-Lasinio (NJL) model ~\cite{Ruggieri:2014bqa,Ferreira:2014kpa,Farias:2016gmy,Zhu:2022irg}, considering the anomalous magnetic moment (AMM) of quarks ~\cite{Andersen:2014xxa,Fayazbakhsh:2014mca,Qiu:2023ezo,Aguirre:2021ljk,Aguirre:2020tiy,Chaudhuri:2019lbw,Chaudhuri:2020lga,Mao:2022dqn,Farias:2021fci,Tavares:2023oln}, or going beyond the mean-field approximation ~\cite{Mao:2016fha}. It is worth noting that the physical origin of the AMM is closely related to chiral symmetry breaking: studies have shown that the AMM can not only be induced by an external magnetic field ~\cite{Ferrer:2013noa,Chang:2010hb,Ferrer:2008dy}, but its dynamic generation mechanism may also be intrinsically linked to the magnetic catalytic effect ~\cite{Mao:2018jdo,Mei:2020jzn,Ferrer:2009nq,Bicudo:1998qb,Ferrer:2008dy,Ferrer:2015wca,Ghosh:2021dlo}. Therefore, once quarks obtain dynamical mass, they should also obtain a dynamical AMM.

In recent years, the study of magnetic field effects on the hadron spectra has induced widespread attention. As Goldstone bosons with chiral (isospin) symmetry breaking, the properties of pseudoscalar mesons under finite temperatures, magnetic fields, and densities have been extensively studied ~\cite{Andersen:2012dz,Kamikado:2013pya,Krein:2021sco,Xu:2020yag,Wang:2017vtn,Coppola:2018vkw,Zhang:2016qrl,Liu:2018zag,Li:2020hlp,Mao:2018dqe,Colucci:2013zoa,Sheng:2020hge,GomezDumm:2017jij,Avancini:2018svs,Li:2022jqf,Kojo:2021gvm,Mishra:2020ckq}. However, existing research has mostly focused on the two-flavor NJL model, and the correlation between AMMs and the mass spectrum of neutral pseudoscalar mesons in the three-flavor system has not been systematically explored, which limits the model's predictive ability for the phase structure of multiflavor quark systems.

Based on the three-flavor NJL model, this article systematically investigates for the first time the influence of quark AMMs on the mass spectra of neutral pseudoscalar mesons ($\pi$, \textit{K}, $\eta$, $\eta^{'}$) under the joint regulation of magnetic field, temperature, and chemical potential. By introducing the AMMs at the quark level, and without loss of generality, two settings of AMMs, namely, AMM1 and AMM2, are introduced. The aim is to reveal the following key issues: (1) how AMM modifies the dependence of quark mass on the magnetic field, thereby reshaping the QCD phase diagram, (2) the enhancement effect of AMM-induced flavor symmetry breaking on meson mixing (such as $\pi_{0}-\eta-\eta^{'}$ mixing), and (3) at finite $\mu$ and $T$, the correlation mechanism between the abrupt change behavior of meson mass and chiral symmetry restoration under strong magnetic fields.  This study not only provides a new perspective for understanding the IMC effect but also verifies the necessity of AMMs in effective models by qualitatively comparing with LQCD results.

The structure of the full text is as follows: Sec. II introduces the theoretical framework of the three-flavor NJL model with AMMs, Sec. III presents numerical results for quark masses, phase diagrams, and meson mass spectra, and analyzes their physical implications, and Sec. IV summarizes the main conclusions and presents discussions.

\section{MODEL AND FORMALISM }\label{sec:2}

The SU(3) NJL model with the AMM of quarks is defined via following the Lagrangian density under an external constant magnetic field,
\begin{equation}\label{eq:1}
	\begin{aligned}
		&\mathcal{L}_{NJL}=\sum_{\textit{f}=u,d,s}\bar{\psi_{\textit{f}}}(\textit{i}\gamma^{\mu}D_{\mu}^{\textit{(\textit{f})}}-m_{\textit{f}}-\frac{1}{2}e_{\textit{f}}\kappa_{\textit{f}}\sigma^{\mu\nu}\textit{F}_{\mu\nu})\psi_{\textit{f}}+\\&\textit{G}\sum_{\textit{a}=0}^{8}[(\bar{\psi}\lambda_{\textit{a}}\psi)^{2}+(\bar{\psi}\textit{i}\gamma^{5}\lambda_{\textit{a}}\psi)^{2}]-\textit{K}(det[\bar{\psi}(1+\gamma^{5})\psi]+det[\bar{\psi}(1-\gamma^{5})\psi]),
	\end{aligned}
\end{equation}
where $\psi=(\psi_{u},\psi_{d},\psi_{s})$ are Dirac spinor for u, d, and s quarks, respectively, $\lambda_{\textit{a}}$ with $\textit{a}=1,2,...,8$ are Gell-Mann matrices with $\lambda_{0}=\sqrt{\dfrac{2}{3}}$, $m_{\textit{f}}$ is the current quark mass, $\textit{D}_{\mu}^{(\textit{f})}=\partial_{\mu}+\textit{i}\textit{q}_{\textit{f}}\textit{A}_{\mu}$ is the covariant derivative with quark charge $\textit{q}_{\textit{f}}=diag(\frac{2}{3}\textit{e},-\frac{1}{3}\textit{e},-\frac{1}{3}\textit{e})$, $\textit{F}_{\mu\nu}=\partial_{\mu}\textit{A}_{\nu}-\partial_{\nu}\textit{A}_{\mu}$ is the electromagnetic field strength, and $\kappa_{\textit{f}}$ denotes the AMM of quarks with flavor $\textit{f}=u,d,s$. Without loss of generality, we consider the magnetic field along $\textit{z}$ direction and take Landau gauge $\textit{A}_{\mu}=(0,0,\textit{xB},0)$. The last term in Eq. (1) is the six-fermion Kobayashi-Maskawa-t$^{'}$ Hooft interaction ~\cite{tHooft:1976snw,tHooft:1986ooh,Kunihiro:1987bb,Reinhardt:1988xu,Bernard:1987sg} that breaks the $\textit{U}_{\textit{A}}(1)$ symmetry.

Under the mean-field approximation, the thermodynamic potential is taken as
\begin{equation}\label{eq:2}
	\begin{aligned}
		\varOmega_{\textit{MF}}=\varOmega_{\textit{q}}+2\textit{G}(\sigma_{u}^{2}+\sigma_{d}^{2}+\sigma_{s}^{2})-4\textit{K}\sigma_{u}\sigma_{d}\sigma_{s},
	\end{aligned}
\end{equation}
where
\begin{equation}\label{eq:3}
	\begin{aligned}
		\varOmega_{\textit{q}}=-3\sum_{\textit{f}=u,d,s}\frac{|\textit{q}_{\textit{f}}\textit{B}|}{2\pi}\sum_{n}\sum_{s=\pm1}\int\frac{dp_{z}}{2\pi}[\textit{E}_{\textit{fns}}
+\textit{T}\ln(1+e^{-\frac{\textit{E}_{\textit{fns}}+\mu}{\textit{T}}})+\textit{T}\ln(1+e^{-\frac{\textit{E}_{\textit{fns}}-\mu}{\textit{T}}})],
	\end{aligned}
\end{equation}
where $\textit{E}_{\textit{fns}}=\sqrt{\textit{p}_{\textit{z}}+(\textit{M}_{\textit{nf}}-\textit{s}\kappa_{\textit{f}}\textit{e}_{\textit{f}}\textit{B})^{2}} , \textit{M}_{\textit{nf}}=\sqrt{\textit{M}_{\textit{f}}^{2}+2\textit{n}|\textit{e}_{\textit{f}}\textit{B}|}$, and chiral condensates $\sigma_{\textit{f}}$. The corresponding effective quark masses satisfy
\begin{equation}\label{eq:4}
	\begin{aligned}
		\textit{M}_{\textit{f}}=\textit{m}_{\textit{f}}-4\textit{G}\sigma_{\textit{f}}+2\textit{K}\prod_{\textit{f'}\ne\textit{f}}\sigma_{\textit{f'}}.
	\end{aligned}
\end{equation}

By minimizing the thermodynamic potential $\partial\varOmega_{\textit{MF}}/\partial\sigma_{\textit{f}}=0$,one can obtain the chiral condensates $\sigma_{\textit{f}}=<\bar{\psi_{\textit{f}}}\psi_{\textit{f}}>$ as
\begin{equation}\label{eq:5}
	\begin{aligned}
		<\bar{\psi_{\textit{f}}}\psi_{\textit{f}}>=-\frac{|\textit{e}_{\textit{f}}\textit{B}|}{(2\pi)^{2}}\int dp_{\textit{z}}\sum_{\textit{n,s}}\frac{\textit{M}_{\textit{f}}}{\textit{E}_{\textit{fns}}}(1-\frac{\textit{s}\kappa_{\textit{f}}\textit{e}_{\textit{f}}\textit{B}}{\textit{M}_{\textit{nf}}})(1-\frac{1}{e^{(\textit{E}_{\textit{fns}}-\mu)/\textit{T}}+1}-\frac{1}{e^{(\textit{E}_{\textit{fns}}+\mu)/\textit{T}}+1}),
	\end{aligned}
\end{equation}
where $\textit{s}=+1$ for $u$ quark, $\textit{s}=-1$ for $d$ and $s$ quarks, respectively, at the lowest Landon level.

To obtain the expression for the meson propagator, it is necessary to introduce the self-consistent Bethe-Salpeter equation with the random phase approximation (RPA) in the NJL model ~\cite{Nambu:1961tp,Nambu:1961fr,Klevansky:1992qe,Volkov:1993jw,Buballa:2003qv,Hatsuda:1994pi}. Under the mean-field approximation, it is necessary to transform the six-fermion interaction into an effective four-fermion interaction ~\cite{Rehberg:1995kh}. The corresponding Lagrangian density can be expressed as
\begin{equation}\label{eq:6}
	\begin{aligned}
		&\mathcal{L}_{NJL}=\sum_{\textit{f}=u,d,s}\bar{\psi_{\textit{f}}}(\textit{i}\gamma^{\mu}D_{\mu}^{\textit{(\textit{f})}}-m_{\textit{f}}-\frac{1}{2}e_{\textit{f}}\kappa_{\textit{f}}\sigma^{\mu\nu}\textit{F}_{\mu\nu})\psi_{\textit{f}}+\\&\sum_{\textit{a}=0}^{8}[\textit{K}_{\textit{a}}^{-}(\bar{\psi}\lambda_{\textit{a}}\psi)^{2}+\textit{K}_{\textit{a}}^{+}(\bar{\psi}\textit{i}\gamma^{5}\lambda_{\textit{a}}\psi)^{2}]+\\&\textit{K}_{30}^{-}(\bar{\psi}\lambda_{3}\psi)(\bar{\psi}\lambda_{0}\psi)+\textit{K}_{30}^{+}(\bar{\psi}\textit{i}\gamma^{5}\lambda_{3}\psi)(\bar{\psi}\textit{i}\gamma^{5}\lambda_{0}\psi)+\textit{K}_{03}^{-}(\bar{\psi}\lambda_{0}\psi)(\bar{\psi}\lambda_{3}\psi)+\textit{K}_{03}^{+}(\bar{\psi}\textit{i}\gamma^{5}\lambda_{0}\psi)(\bar{\psi}\textit{i}\gamma^{5}\lambda_{3}\psi)+\\&\textit{K}_{80}^{-}(\bar{\psi}\lambda_{8}\psi)(\bar{\psi}\lambda_{0}\psi)+\textit{K}_{80}^{+}(\bar{\psi}\textit{i}\gamma^{5}\lambda_{8}\psi)(\bar{\psi}\textit{i}\gamma^{5}\lambda_{0}\psi)+\textit{K}_{08}^{-}(\bar{\psi}\lambda_{0}\psi)(\bar{\psi}\lambda_{8}\psi)+\textit{K}_{08}^{+}(\bar{\psi}\textit{i}\gamma^{5}\lambda_{0}\psi)(\bar{\psi}\textit{i}\gamma^{5}\lambda_{8}\psi)+\\&\textit{K}_{83}^{-}(\bar{\psi}\lambda_{8}\psi)(\bar{\psi}\lambda_{3}\psi)+\textit{K}_{83}^{+}(\bar{\psi}\textit{i}\gamma^{5}\lambda_{8}\psi)(\bar{\psi}\textit{i}\gamma^{5}\lambda_{3}\psi)+\textit{K}_{38}^{-}(\bar{\psi}\lambda_{3}\psi)(\bar{\psi}\lambda_{8}\psi)+\textit{K}_{38}^{+}(\bar{\psi}\textit{i}\gamma^{5}\lambda_{3}\psi)(\bar{\psi}\textit{i}\gamma^{5}\lambda_{8}\psi),
	\end{aligned}
\end{equation}
where the effective coupling constants are taken as
\begin{equation}\label{eq:7}
	\begin{aligned}
		&\textit{K}_{0}^{\pm}=\textit{G}\pm\frac{1}{3}\textit{K}(\sigma_{\textit{u}}+\sigma_{\textit{d}}+\sigma_{\textit{s}}),\\&\textit{K}_{1}^{\pm}=\textit{K}_{2}^{\pm}=\textit{K}_{3}^{\pm}=\textit{G}\pm\dfrac{1}{2}\textit{K}\sigma_{\textit{s}},\\&\textit{K}_{4}^{\pm}=\textit{K}_{5}^{\pm}=\textit{G}\pm\dfrac{1}{2}\textit{K}\sigma_{\textit{d}},\\&\textit{K}_{6}^{\pm}=\textit{K}_{7}^{\pm}=\textit{G}\pm\dfrac{1}{2}\textit{K}\sigma_{\textit{u}},\\&\textit{K}_{8}^{\pm}=\textit{G}\pm\dfrac{1}{6}\textit{K}(2\sigma_{\textit{u}}+2\sigma_{\textit{d}}-\sigma_{\textit{s}}),\\&\textit{K}_{30}^{\pm}=\textit{K}_{03}^{\pm}=\mp\frac{1}{2\sqrt{6}}\textit{K}(\sigma_{\textit{u}}-\sigma_{\textit{d}}),\\&\textit{K}_{80}^{\pm}=\textit{K}_{08}^{\pm}=\pm\frac{\sqrt{2}}{12}\textit{K}(\sigma_{\textit{u}}+\sigma_{\textit{d}}-2\sigma_{\textit{s}}),\\&\textit{K}_{83}^{\pm}=\textit{K}_{38}^{\pm}=\pm\frac{1}{2\sqrt{3}}\textit{K}(\sigma_{\textit{u}}-\sigma_{\textit{d}}),
	\end{aligned}
\end{equation}
and chiral condensates are obtained as
\begin{equation}\label{eq:8}
	\begin{aligned}
		\sigma_{\textit{u}}=<\bar{\psi_{\textit{u}}}\psi_{\textit{u}}>,\sigma_{\textit{d}}=<\bar{\psi_{\textit{d}}}\psi_{\textit{d}}>,\sigma_{\textit{s}}=<\bar{\psi_{\textit{s}}}\psi_{\textit{s}}>.
	\end{aligned}
\end{equation}

We only consider neutral mesons, whose quark propagator Schwinger phase can be cancelled. For mesons that do not mix with other mesons, their propagators can be represented by polarization loops ~\cite{Klevansky:1992qe} through RPA as
\begin{equation}\label{eq:9}
	\begin{aligned}
		\frac{1}{\textit{i}}[\Pi_{\textit{ps}}(\textit{p})]_{\textit{ij}}=-\textit{N}_{\textit{c}}\sum_{\textit{f}\textit{f'}}\int\frac{\textit{d}^{4}\textit{k}}{(2\pi)^{4}}
tr[\gamma^{5}(\textit{T}_{\textit{i}})_{\textit{f}\textit{f}}\textit{S}^{\textit{f}}(\textit{p+k})\gamma^{5}(\textit{T}_{\textit{j}})_{\textit{f}\textit{f'}}
\textit{S}^{\textit{f'}}(\textit{k})],
	\end{aligned}
\end{equation}
where the flavor indices $\textit{f}$ and $\textit{f'}$ are explicitly included and the trace corresponds to the spinor trace only, $\textit{S}^{\textit{f}}$ is the quark propagator of flavor $\textit{f}$,  and the vertex of nonmixing mesons can be given as
\begin{equation}\label{eq:10}
	\begin{aligned}
		\textit{T}_{\textit{i}}=\begin{cases}
			\frac{1}{\sqrt{2}} (\lambda_1 \pm i\lambda_2), & \text{for $\pi^\pm$} \\
			\frac{1}{\sqrt{2}} (\lambda_6 \pm i\lambda_7), & \text{for $K^0, \bar{K}^0$} \\
			\frac{1}{\sqrt{2}} (\lambda_4 \pm i\lambda_5), & \text{for $K^\pm$}
		\end{cases}.
		\,
	\end{aligned}
\end{equation}

Therefore, the $\textit{K}^{0}$ meson propagator can be given as
\begin{equation}\label{eq:11}
	\begin{aligned}
		\textit{M} = \frac{2K_6^+}{1 - 2K_6^+ \Pi_{\textit{K}^{0}}^{\textit{ps}}(p)},
	\end{aligned}
\end{equation}
and the mass is determined by solving the equation
\begin{equation}\label{eq:12}
	\begin{aligned}
		1 - 2 \textit{K}_6^+ \Pi_{{\textit{K}}^0}^{\textit{ps}}(\textit{p}_0, \vec{\textit{p}} = 0) = 0.
	\end{aligned}
\end{equation}

The magnetic field can break isospin symmetry, and the AMM of quarks can even enhance this breaking effect, leading to the production of mixing states $\pi_{0}-\eta-\eta^{'}$. For the mixing mesons, the meson propagator~\cite{Costa:2005cz} can be expressed as
\begin{equation}\label{eq:13}
	M = \begin{bmatrix}
		M_{00} & M_{03} & M_{08} \\
		M_{30} & M_{33} & M_{38} \\
		M_{80} & M_{83} & M_{88}
	\end{bmatrix} = 2K^+\left(1 - 2{\Pi}^{\textit{ps}}K^+\right)^{-1},
\end{equation}
where
\begin{equation}\label{eq:14}
	K^+ =
	\begin{bmatrix}
		K_{00}^+ & K_{03}^+ & K_{08}^+ \\
		K_{30}^+ & K_{33}^+ & K_{38}^+ \\
		K_{80}^+ & K_{83}^+ & K_{88}^+
	\end{bmatrix},
\end{equation}
and
\begin{equation}\label{eq:15}
	{\Pi}^{\textit{ps}} =
	\begin{bmatrix}
		{\Pi}_{00}^{\textit{ps}} & {\Pi}_{03}^{\textit{ps}} & {\Pi}_{08}^{\textit{ps}} \\
		{\Pi}_{30}^{\textit{ps}} & {\Pi}_{33}^{\textit{ps}} & {\Pi}_{38}^{\textit{ps}} \\
		{\Pi}_{80}^{\textit{ps}} & {\Pi}_{83}^{\textit{ps}} & {\Pi}_{88}^{\textit{ps}}
	\end{bmatrix},
\end{equation}
and they satisfy the following equations$K_{03}^+=K_{30}^+,K_{08}^+=K_{80}^+,K_{38}^+=K_{83}^+$, and
\begin{equation}\label{eq:16}
	\begin{aligned}
		\Pi_{00}^{ps} &= \frac{2}{3} \left( \Pi_{u\bar{u}}^{ps} + \Pi_{d\bar{d}}^{ps} + \Pi_{s\bar{s}}^{ps} \right), \\
		\Pi_{03}^{ps} &= \Pi_{30}^{ps} = \sqrt{\frac{2}{3}} \left( \Pi_{u\bar{u}}^{ps} - \Pi_{d\bar{d}}^{ps} \right), \\
		\Pi_{08}^{ps} &= \Pi_{80}^{ps} = \frac{\sqrt{2}}{3} \left( \Pi_{u\bar{u}}^{ps} + \Pi_{d\bar{d}}^{ps} - 2\Pi_{s\bar{s}}^{ps} \right), \\
		\Pi_{33}^{ps} &= \Pi_{u\bar{u}}^{ps} + \Pi_{d\bar{d}}^{ps}, \\
		\Pi_{38}^{ps} &= \Pi_{83}^{ps} = \frac{\sqrt{3}}{3} \left( \Pi_{u\bar{u}}^{ps} - \Pi_{d\bar{d}}^{ps} \right), \\
		\Pi_{88}^{ps} &= \frac{1}{3} \left( \Pi_{u\bar{u}}^{ps} + \Pi_{d\bar{d}}^{ps} + 4\Pi_{s\bar{s}}^{ps} \right).
	\end{aligned}
\end{equation}

The $\pi_{0},\eta,\eta^{'}$ meson masses can be determined by solving the equation as
\begin{equation}\label{eq:17}
	\begin{aligned}
		\det\left[M^{-1}(p_0, \vec{\textit{p}} = 0)\right] = 0.
	\end{aligned}
\end{equation}

One can simplify the inverse of meson propagator matrix $M$ as
\begin{equation}\label{eq:18}
	M^{-1} = \frac{1}{2 \det K^+} \begin{pmatrix}
		D & A & B \\
		A & E & C \\
		B & C & F
	\end{pmatrix},
\end{equation}
with
\begin{equation}\label{eq:19}
	\begin{aligned}
		A &= K^+_{08} K^+_{38} - K^+_{03} K^+_{88} - 2\Pi^{ps}_{03} \det K^+, \\
		B &= K^+_{08} K^+_{38} - K^+_{08} K^+_{33} - 2\Pi^{ps}_{08} \det K^+, \\
		C &= K^+_{08} K^+_{03} - K^+_{00} K^+_{38} - 2\Pi^{ps}_{38} \det K^+, \\
		D &= K^+_{33} K^+_{38} - (K^+_{38})^2 - 2\Pi^{ps}_{00} \det K^+, \\
		E &= K^+_{00} K^+_{88} - (K^+_{08})^2 - 2\Pi^{ps}_{33} \det K^+, \\
		F &= K^+_{00} K^+_{33} - (K^+_{03})^2 - 2\Pi^{ps}_{88} \det K^+.
	\end{aligned}
\end{equation}

It is worth emphasizing that we can obtain the meson mass through Eq. (17), where the three pairs of roots correspond to the masses of $\pi_{0},\eta,\eta'$ mesons. To calculate the polarization loop, we need to use the quark propagator. We can first solve the Dirac equation corresponding to the first term in Eq. (1). Then, the quark propagator is obtained by employing the standard canonical quantization in quantum field theory and the real-time formula of thermal field theory. When $n=0$, the quark propagator of the flavor $f$ in momentum space with the real-time formula can be obtained as ~\cite{Aguirre:2016zqw}
\begin{equation}\label{eq:20}
	\begin{aligned}
		&S_{f0}(k) = e^{-\frac{{{k}_\perp}^2}{|e_f B|}} \left[ (k_0 \gamma^0 - k_z \gamma^3) + (M_f - s \kappa_f e_f B) \right] \\&\times(1 + \text{sign}(e_f) i \gamma^1 \gamma^2) \left\{ \frac{1}{k_0^2 - {E_{f0s}}^2 + i \varepsilon} + 2 \pi i n_F(k_0) \delta(k_0^2 - {E_{f0s}}^2) \right\},
	\end{aligned}
\end{equation}
and when $n\geq1$, the quark propagator of the flavor $f$ in momentum space can be given as
\begin{equation}\label{eq:21}
	\begin{aligned}
		&S_f(k) = e^{-\frac{{{k}_\perp}^2}{|e_f B|}} \sum_{n,s} (-1)^n \frac{1}{2M_{nf}} \{ (M_{nf} + sM_f) [(k_0 \gamma^0 - k_z \gamma^3) + (sM_{nf} - \kappa_f e_f B)] \\&\times
		(1 + \text{sign}(e_f) i \gamma^1 \gamma^2) L_n \left( \frac{2{{k}_\perp}^2}{|e_f B|} \right) - (M_{nf} - sM_f) [(k_0 \gamma^0 - k_z \gamma^3) \\&
		- (sM_{nf} - \kappa_f e_f B)] (1 - \text{sign}(e_f) i \gamma^1 \gamma^2) L_{n-1} \left( \frac{2{{k}_\perp}^2}{|e_f B|} \right) \\&
		- 4[(k_0 \gamma^0 - k_z \gamma^3) \text{sign}(e_f) i \gamma^1 \gamma^2 + (sM_{nf} - \kappa_f e_f B)] (-k_x \gamma^1 - k_y \gamma^2) s L_{n-1}^1 \left( \frac{2{{k}_\perp}^2}{|e_f B|} \right) \} \\&\times
		\left\{ \frac{1}{k_0^2 - {E_{f0s}}^2 + i \varepsilon} + 2 \pi i n_F(k_0) \delta(k_0^2 - {E_{f0s}}^2) \right\},
	\end{aligned}
\end{equation}
where $E_{fns}$ represents the energy eigenvalue, $L_{n}$ stands for the Laguerre polynomial of order $n$, the index $s=\pm1$ denotes the spin quantum number, and $s=+1$ for $u$ quark and $s=-1$ for $d$ and $s$ quarks at the lowest Landau level. $n_F(k_0)$ that appears in Eqs. (20) and (21) can be obtained as
\begin{equation}\label{eq:22}
	\begin{aligned}
		n_F(k_0) = \theta(k_0) \frac{1}{e^{(k_0 - \mu)/T} + 1} + \theta(-k_0) \frac{1}{e^{-(k_0 - \mu)/T} + 1}.
	\end{aligned}
\end{equation}

For the kaon meson, the polarization loop can be obtained as
\begin{equation}\label{eq:23}
	\begin{aligned}
		&\Pi^{ps}_{K^0}(p_0 = m_{K^0}) = \frac{\beta N_c}{2(2\pi)^2} \sum_{n,s,l} (1 + sl \frac{M_d M_s + 2\beta n}{M_{nd} M_{ns}}) \\&\times
		\left\{ \int dk_z \frac{1}{E_{dns}} \left[1- \frac{1}{e^{(E_{dns} - \mu)/T} + 1} - \frac{1}{e^{(E_{dns} + \mu)/T} + 1} \right] \right. \\&
		+ \left. \int dk_z \frac{1}{E_{snl}} \left[1- \frac{1}{e^{(E_{snl} - \mu)/T} + 1} - \frac{1}{e^{(E_{snl} + \mu)/T} + 1} \right] \right. \\&
		+ \left. \left\{ [(M_{nd} - s \kappa_d e_d B) - sl(M_{ns} - l \kappa_s e_s B)]^2 - p_0^2 \right\} B(m_d, m_s) \right\},
	\end{aligned}
\end{equation}
where
\begin{equation}\label{eq:24}
	\begin{aligned}
		&B(m_d, m_s) =\\& \int dk_z \left\{ \frac{1}{E_{dns}} \left[ \frac{1}{(E_{dns} + p_0)^2 - {E_{snl}}^2} \frac{1}{e^{-(E_{dns} + \mu)/T} + 1} - \frac{1}{(E_{dns} - p_0)^2 - {E_{snl}}^2} \frac{1}{e^{(E_{dns} - \mu)/T} + 1} \right] \right. \\&
		+ \left. \frac{1}{E_{snl}} \left[ \frac{1}{(E_{snl} - p_0)^2 - {E_{dns}}^2} \frac{1}{e^{-(E_{snl} + \mu)/T} + 1} - \frac{1}{(E_{snl} + p_0)^2 - {E_{dns}}^2} \frac{1}{e^{(E_{snl} - \mu)/T} + 1} \right] \right\}.
	\end{aligned}
\end{equation}

For $u$, $d$ and $s$ quarks, the polarization loops are obtained as
\begin{equation}\label{eq:25}
	\begin{aligned}
		&\Pi^{ps}_{f\bar{f}}(p_0 = m) = \frac{\beta N_c}{(2\pi)^2} \sum_{n,s} \int dk_z \frac{1}{E_{fns}} \left[ 1 - \frac{1}{e^{(E_{fns} - \mu)/T} + 1} - \frac{1}{e^{(E_{fns} + \mu)/T} + 1} \right] \\&
		+ p_0^2 \frac{\beta N_c}{(2\pi)^2} \sum_{n,s} \int dk_z \frac{1}{E_{fns}} \frac{1}{4E_{fns}^2 - p_0^2} \left[ 1 - \frac{1}{e^{(E_{fns} - \mu)/T} + 1} - \frac{1}{e^{(E_{fns} + \mu)/T} + 1} \right],
	\end{aligned}
\end{equation}
and the detailed derivation of polarization loops is given in Appendix A.

\section{RESULTS AND ANALYSIS}\label{sec:4}
Since the NJL model is nonrenormalizable, a proper regularization scheme is needed. In our work, we apply the gauge-invariant Pauli-Villars(PV) regularization~\cite{Pauli:1949zm} in the article, which can guarantee the law of causality and effectively avoid the unphysical oscillation at the finite magnetic field. Other schemes, like the proper-time regularization scheme and (improved) magnetic-field- independent regularization (MFIR)~\cite{Avancini:2018svs,Avancini:2019wed,Aguirre:2020tiy,Aguirre:2021ljk}, which separates the vacuum and magnetic contribution, are also effective when dealing with system under a uniform magnetic field. In the case of finite magnetic field considering the nonzero values of the AMM of the quarks, the energy of quarks is given by $E_{fns} = p_z^2 + (M_{nf} - s k_f e_f B)^2 = p_z^2 + M_{eff}^2, \quad M_{nf} = \sqrt{M_f^2 + 2n|e_f B|}$. Therefore, under PV scheme we make the following replacement~\cite{Mao:2016fha,Mao:2016lsr,Mei:2020jzn,Chaudhuri:2021skc,Bao:2024glw}:
\begin{equation}\label{eq:26}
	\begin{aligned}
		\sum_{n=-\infty}^{+\infty} \int_{-\infty}^{+\infty} \frac{dp_z}{2\pi} f(E_{fns}) \rightarrow \sum_{i=0}^{N} c_i \sum_{n=-\infty}^{+\infty} \int_{-\infty}^{+\infty} \frac{dp_z}{2\pi} f(E_{fns,i}),
	\end{aligned}
\end{equation}
where $E_{fns,i} = p_z^2 + M_{eff}^2 + b_i \Lambda^2$ and $c_{0}=1,c_{1}=-3,c_{2}=3,c_{3}=-1$; $b_{0}=0,b_{1}=1,b_{2}=2,b_{3}=3$.

In this paper, we adopt the widely used parameter set~\cite{Carignano:2019ivp,Sheng:2022ssp}, which has been fitted to the vacuum values of the $\pi$ meson decay constant and the masses of the $\pi$ meson, $K$ meson, and $\eta'$ meson, while fixing the constituent light quark mass in vacuum to a value of ~300 MeV at zero temperature. The parameter set is given as $m_{u,d}$ = 10.3 MeV, $m_{s}$ = 236.9 MeV, $\Lambda$ = 781.2 MeV, $G\Lambda^{2}$ = 4.90, and $K\Lambda^{5}$ = 129.8. To simplify the discussion, we adopt constant $\kappa_{f}$ values. Through the Foldy- Wouthuysen transformationn~\cite{Foldy:1949wa}, one can show that the quark magnetic moment is modified as $\mu_f = \left(1 + M_f \kappa_f\right) \frac{e_f}{2 M_f}$, where $M_f$ is the effective quark masses. By fitting the magnetic moments of valence quarks [$\mu_u = (2.08 \pm 0.07) \mu_N; \mu_d = (-1.31 \pm 0.06) \mu_N; \mu_s = (-0.77 \pm 0.06) \mu_N$] obtained within the Geffen-Wilson model~\cite{Dothan:1981ex} to experimental values of the baryonic magnetic moments, one derives the following set of AMMs, ~$\kappa_{u}$ = 0.123, $\kappa_{d}$ = 0.555, and $\kappa_{s}$ = 0.329, ~in units of the nuclear magneton. These larger $\kappa_{f}$  values were employed in Refs.~\cite{Sheng:2022ssp,Qiu:2023kwv} and we take this AMM set as AMM2 in our work. We also selected a second AMM set of smaller $\kappa_{f}$  values: $\kappa_{u}$ = 0.074, $\kappa_{d}$ = 0.127, and $\kappa_{s}$ = 0.053, labeled AMM1 in our work. These values were derived within the constituent quark model framework by first obtaining the magnetic moments $\mu_f$ , then fitting them to experimental values of the baryonic magnetic moments together with effective quark masses estimated within the same framework. This second AMM set has been briefly described and utilized in Ref. \cite{Aguirre:2020tiy} for related studies. The label AMM0 in our article is used for the results of this work when the AMM is not considered.
\begin{figure}[H]
	\centering
	\includegraphics[width=1.05\textwidth]{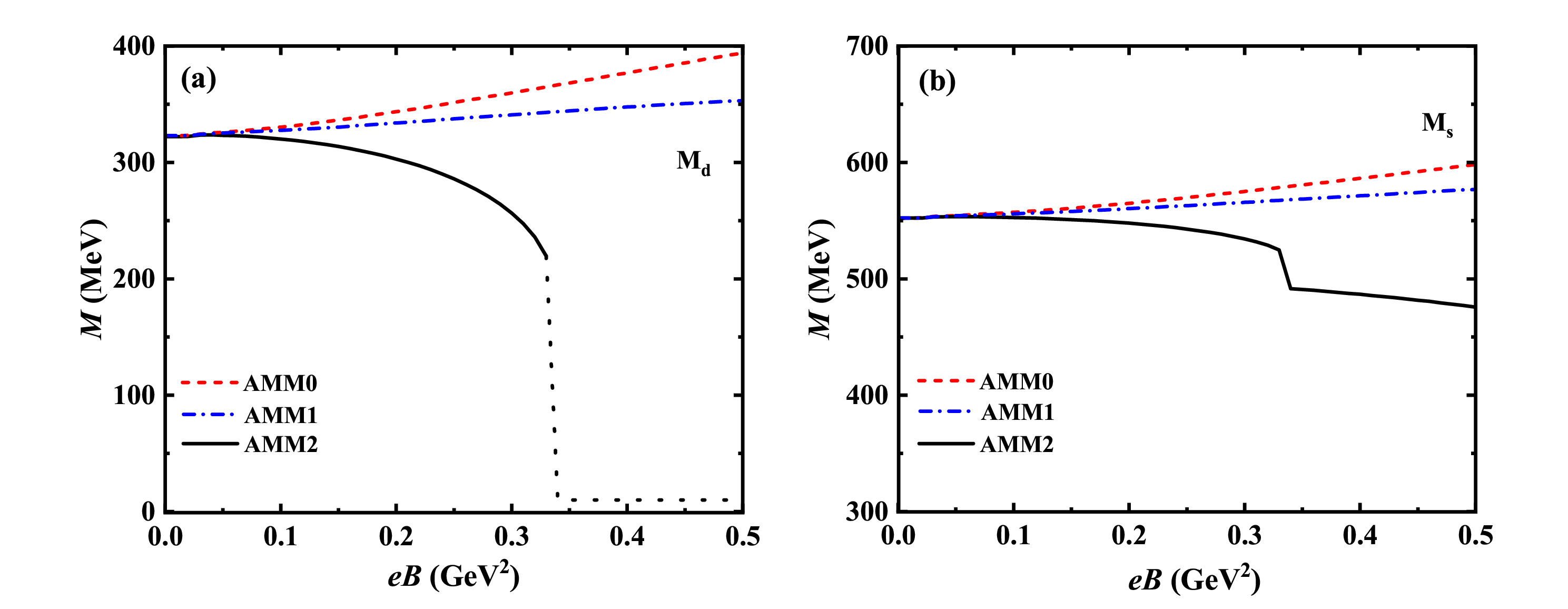}
	\caption{\label{Fig1} The dependence of dynamical quark mass ($M$) on the magnetic field ($eB$) with and without values of AMM at zero temperature. (a) and (b) are for $d$ quark and $s$ quark, respectively.}
	\label{Fig:1}
\end{figure}

Figure~\ref{Fig:1} illustrates the dependence of the dynamical quark masses of $d$ quarks and $s$ quarks on the magnetic field at temperature $T = 0$.  It is
worth noting that the AMM sets in Fig.1 correspond to three different settings, which are AMM0, AMM1, and AMM2, respectively. AMM0 means that the AMM is not considered, that is, all $\kappa$ values are set to zero. AMM1 and AMM2 sets have been mentioned above. This AMM1 set corresponds to smaller $\kappa_{f}$ values (with smaller AMMs), and the AMM2 set corresponds to larger $\kappa_{f}$ values (with larger AMMs).

The study finds that the AMM0 ($\kappa_{f}$ = 0) and  AMM1 (smaller $\kappa_{f}$ ) sets, the masses of both $d$ quarks and $s$ quarks increase with the enhancement of the magnetic field at $T = 0$.  However, when a larger $\kappa_{f}$  value (AMM2 set) is taken into account, a novel phenomenon is observed, where the quark masses exhibit nonmonotonic behavior with increasing magnetic field strength for AMM2. It is observed that their masses increase with the enhancement of the magnetic field for $d$ quarks. But when the magnetic field strength reaches $eB$ = 0.04 GeV$^{2}$, their masses begin to decrease with the increase of the magnetic field, which is also shown by the IMC feature. When the magnetic field strength increases to $eB$ = 0.33 GeV$^{2}$,  the dynamical mass of $d$ quarks drops sharply, directly to its current quark mass, indicating a first-order phase transition in the mass of light quarks because the thermodynamic potential has more than one minimum point.  For the $s$ quark(AMM2 set), its dynamical mass also exhibits nonmonotonic behavior with the enhancement of the magnetic field, starting to decrease when the magnetic field strength increases to $eB$ = 0.06 GeV$^{2}$. Within the entire range of magnetic fields considered, no first-order phase transition occurs for $s$ quarks.

According to LQCD, IMC manifests in two aspects~\cite{Bali:2012zg,Bali:2011qj,DElia:2018xwo}. First, the phase transition temperature decreases with increasing magnetic field strength. Second, the light quark chiral condensate exhibits nonmonotonic or decreasing behavior with increasing magnetic field strength near the phase transition temperature, differing from its behavior at $T$ = 0, where MC occurs. However, this suppression of the chiral condensate and the accompanied reduction of the transition temperature $T_c$ are not predicted by initial studies based on effective models~\cite{Shovkovy:2012zn}. Subsequently, various authors proposed different approaches within the framework of effective models to reproduce this phenomenon-one of which is the incorporation of the quark AMM, as adopted in this work. To specifically illustrate the IMC phenomenon induced by quark AMMs, we performed numerical calculations of the transition temperature $T_c$.

\begin{figure}[H]
	\centering
	\includegraphics[width=1.15\textwidth]{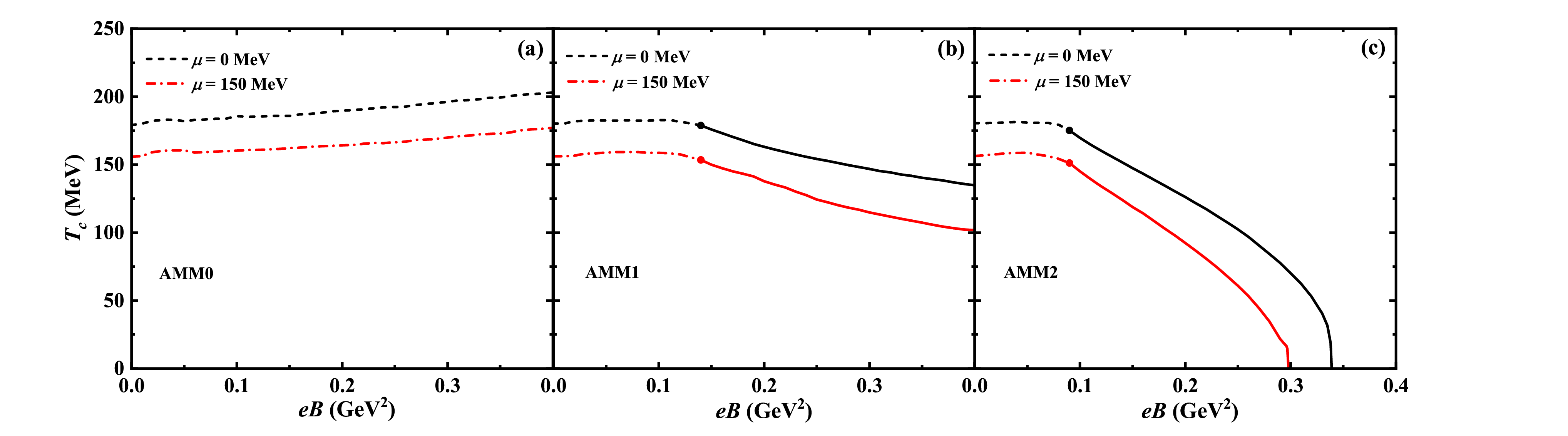}
	\caption{\label{Fig2} The critical temperatures ($T_{c}$) for chiral phase transition as a function of the magnetic field for different chemical potentials without AMM in panel (a), with AMM1 on panel (b), and with AMM2 in panel (c), respectively. The solid lines correspond to chiral first-order transition, the full dots correspond to CEPs, and the other lines correspond to chiral crossover phase transition.}
	\label{Fig:2}
\end{figure}

 As can be seen from Fig.~\ref{Fig:2}(a), when the AMM is not considered, the phase transition temperature increases with the increase of the magnetic field, exhibiting MC for different chemical potentials. However, the situation becomes completely different when the AMM is introduced. As shown in Fig.~\ref{Fig:2}(b) and Fig.~\ref{Fig:2}(c), it can be observed that the AMM has an inhibitory effect on the phase transition temperature $T_{c}$, exhibiting IMC characteristics. Furthermore, we have also discovered that for different chemical potentials in the $(T_{c}-eB)$ phase diagram, there exists critical endpoints (CEPs) of phase transition. The CEP in the phase transition refers to the point where the first-order phase boundary and the crossover phase boundary intersect on the QCD phase diagram. This point marks the transition of the phase transition nature, from a first-order phase transition (discontinuous physical quantities) to a crossover phase transition (continuous physical quantity changes but with a critical point). We also discussed the influence of AMMs on the $(T_{c}-eB)$ phase diagram and found that except in the weak magnetic field region ($eB <$ 0.08 GeV$^{2}$ for AMM1 and $eB <$  0.12 GeV$^{2}$ for AMM2), the critical temperature of the chiral phase transition decreases with the increasing magnetic field, indicating the phenomenon of IMC.

Notably, the chiral phase transition exhibits crossover behavior in weak magnetic fields, regardless of zero or finite chemical potential, but evolves into a first-order transition in strong magnetic fields. For the AMM1 set, when $\mu$ = 0 MeV, the CEP is located at $eB$ = 0.14 GeV$^{2}$ and $T$ = 178.74 MeV, whereas when $\mu$ = 150 MeV, the CEP shifts to $eB$ = 0.14 GeV$^{2}$ and $T$ = 153.5 MeV. For the AMM2 set, when $\mu$ = 0 MeV, the CEP  is situated at $eB$ = 0.09 GeV$^{2}$ and $T$ = 175.1 MeV, whereas when $\mu$ = 150 MeV, the CEP moves to $eB$ = 0.09 GeV$^{2}$ and $T$ = 151.2 MeV.

It is worth noting that, as shown in Fig.~\ref{Fig:1}(a), the $d$ quark mass calculated with the AMM1 does not exhibit a first-order phase transition under the considered magnetic fields, whereas a first-order phase transition appears in Fig.~\ref{Fig:2}(b). This difference is readily understood: Fig.~\ref{Fig:1} presents the quark mass evaluated at zero temperature, thereby reflecting how the strength of chiral symmetry breaking evolves with the magnetic field, whereas Fig.~\ref{Fig:2} investigates the magnetic field dependence of the critical temperature $T_{c}$ at finite temperature. Previous systematic studies on quark masses at finite temperature have shown that the magnetic field required to trigger a first-order transition becomes weak as the temperature increases~\cite{Fayazbakhsh:2014mca}. Therefore, it can be concluded that although no first-order phase transition occurs for light quark masses with small AMM1 at zero temperature, it is expected that at finite temperature, the mass may undergo a first-order phase transition within the considered magnetic field, consistent with the first-order transition claimed in Fig.~\ref{Fig:2}.

\begin{figure}[H]
	\centering
	\includegraphics[width=1.05\textwidth]{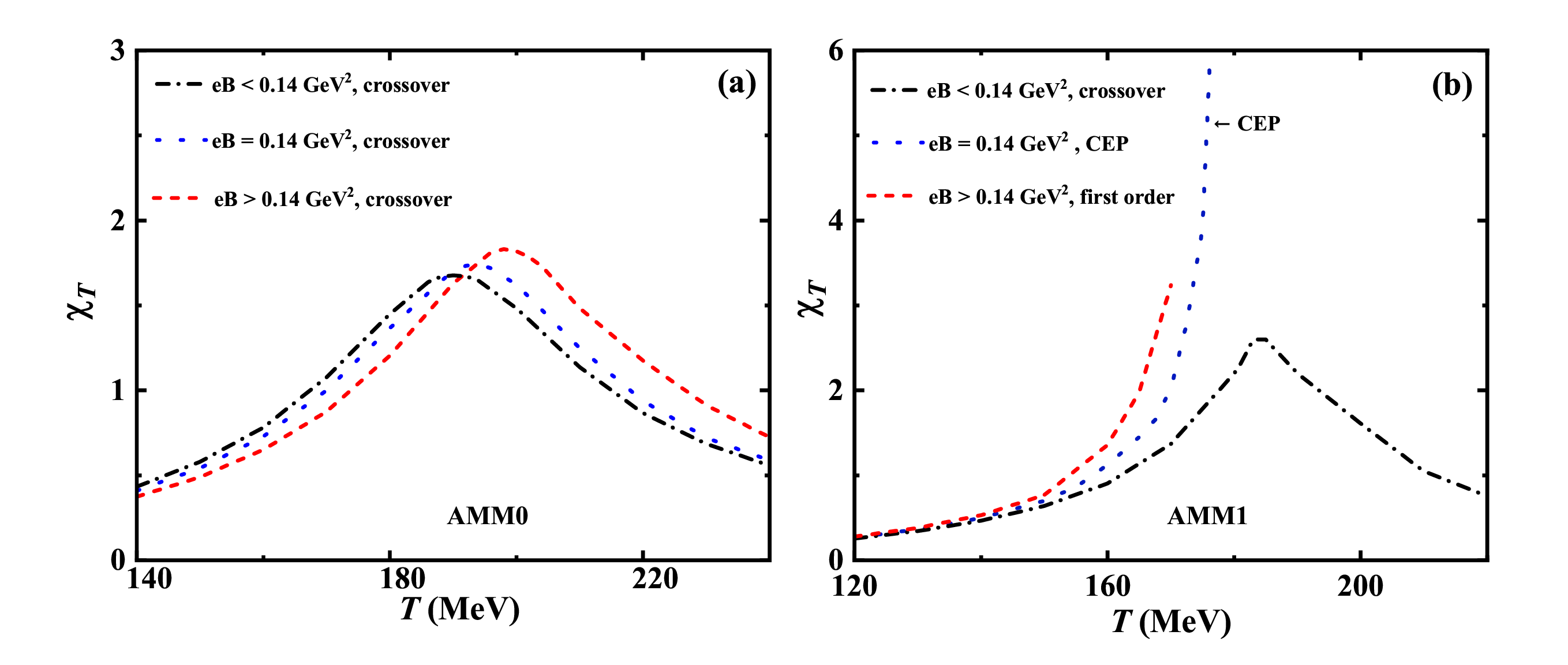}
	\caption{\label{Fig3} The behavior of the thermal susceptibility $\chi_{T}$ in different orders of phase transitions for AMM0 and AMM1 at zero chemical potential.}
	\label{Fig:3}
\end{figure}

Taking the cases of AMM0 and AMM1 at zero chemical potential as examples (other cases being similar), Fig.~\ref{Fig:3} shows the behavior of the thermal susceptibility $\chi_{T}$ in different orders of phase transitions, $\chi_T = \frac{\partial \langle \psi_f \overline{\psi}_f \rangle}{\partial T}$ ~\cite{Lu:2015naa}. Since the chiral condensate $\langle \psi_f \overline{\psi}_f \rangle$ has dimensions of $MeV^{3}$, we employ a dimensionless form $\chi_T = \frac{\partial \langle \psi_f \overline{\psi}_f \rangle^{1/3}}{\partial T}$ for numerical convenience, which does not alter the physical conclusions. Other types of susceptibility can be found in Ref.~\cite{Lu:2015naa}. Considering AMM1, when the magnetic field increases to a certain value (like 0.14 $GeV^{2}$ for AMM1), the $\Omega$ first begins to have two local minimum points and a CEP appears. $\chi_{T}$ diverge at the CEP. Along the first-order phase transition line, the system's thermodynamic potential $\Omega$ exhibits two local minimum points and the thermal susceptibility $\chi_{T}$ undergoes a discontinuous change. In the crossover phase, the thermal susceptibility $\chi_{T}$ remains strictly continuous. When AMM1 is neglected, the NJL system remains in the crossover phase throughout the entire range of the magnetic field, and the thermal susceptibility $\chi_{T}$ always remains continuous. Within the crossover phase, the thermodynamic potential possesses only a single global minimum point. The behavior of the susceptibility, multisolution in $\Omega$, and the order of phase transitions have been comprehensively analyzed in Refs.~\cite{Lu:2015naa,Cui:2013aba,Cui:2018bor}.
At high density, where a first phase transition occurs, two local minimum points of $\Omega$ remain $\sigma_f$ finite without considering AMMs. Unlike the case without considering AMM, the first-order transition is strikingly pronounced and drastic when considering the AMMs: one minimum point remains $\sigma_f$ finite, whereas the other drives $\sigma_f$ close to zero. Due to this reason, the thermal susceptibility $\chi_{T}$ manifests two distinct characteristics. First, at a first-order phase transition, thermal susceptibility $\chi_{T}$ increases monotonically with temperature until it reaches a specific value and then disappears. Nevertheless, in this case, one can also state that the thermal susceptibility $\chi_{T}$ exhibits a discontinuous change. Second, at the CEP, the thermal susceptibility $\chi_{T}$ remains divergent. Similarly, the thermal susceptibility $\chi_{T}$ also disappears for temperatures above $T > T_{CEP}$. This behavior differs from the scenario observed at high chemical potential without the AMM~\cite{Costa:2008gr}.
\begin{figure}[H]
	\centering
	\includegraphics[width=1.05\textwidth]{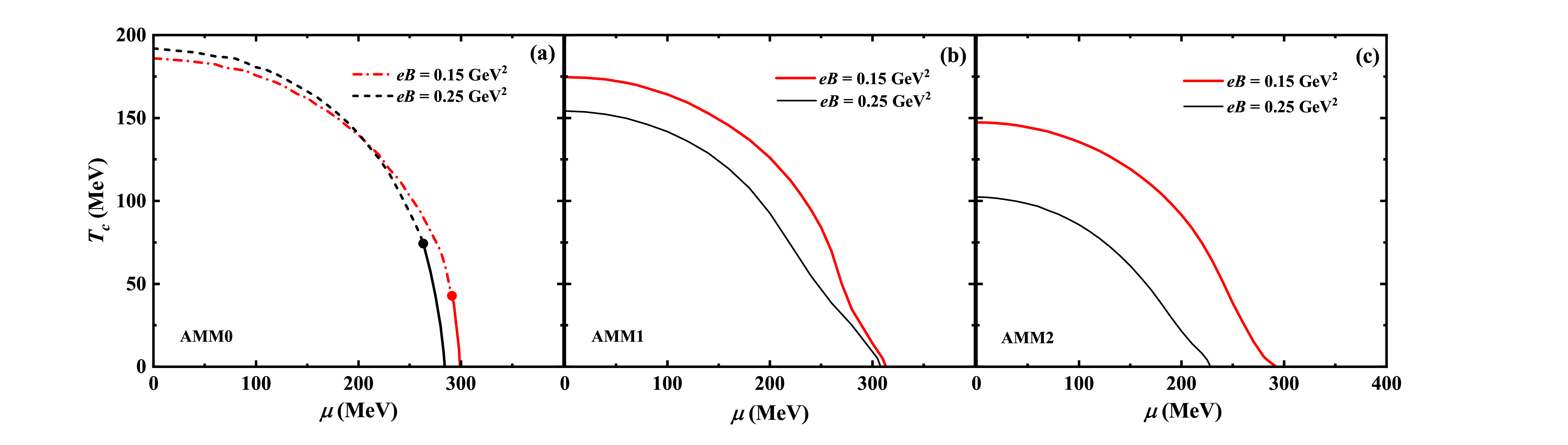}
	\caption{\label{Fig4} The critical temperatures ($T_{\textrm{c}}$) for chiral phase transition as a function of chemical potential for different magnetic fields. Panel (a) is for that without considering quark AMM, and panels (b) and (c) are for considering quark AMM1 and AMM2. The solid lines correspond to chiral first-order transition, the full dots correspond to CEPs, and the other lines correspond to chiral crossover phase transition.}
	\label{Fig:4}
\end{figure}
Figure ~\ref{Fig:4} illustrates the $(T_{\textrm{c}}-\mu)$ phase diagram with different magnetic fields ($eB$ = 0.15 GeV$^{2}$ and 0.25 GeV$^{2}$) without AMM (a), with AMM1 (b), and with AMM2 (c). Solid and dashed lines represent first-order and crossover phase transitions, respectively, with CEPs clearly marked.  Figures~\ref{Fig:4}(b) and ~\ref{Fig:4}(c) show that with the introduction of the quark AMM, the $(T_{\textrm{c}}-\mu)$ phase diagram clearly exhibits first-order phase transition for the magnetic fields of $eB$ = 0.15 GeV$^{2}$ and 0.25 GeV$^{2}$, respectively. This is in contrast to the case shown in Fig.~\ref{Fig:4}(a), where AMM is not considered, where the crossover phase transitions exist in the high- temperature and low chemical potential region, and the first-order transitions exist in the low temperature and high chemical potential region. The CEPs without the AMM, are located at $\mu_{\textrm{CEP}}\approx$ 291.2 MeV, $T_{\textrm{CEP}}$ = 42.6 MeV for $eB$ = 0.15 GeV$^{2}$, and at $\mu_{\textrm{CEP}}\approx$ 263.3 MeV, $T_{\textrm{CEP}}$ = 74.1 MeV for $eB$ = 0.25 GeV$^{2}$.

Figures~\ref{Fig:4}(b) and~\ref{Fig:4}(c) show that the introduction of the AMM  suppresses chiral symmetry breaking, reduces the critical temperature $T_{\textrm{c}}$, and destabilizes the crossover phase transition. Due to the coupling between AMMs and the magnetic field, the constituent quark mass decreases sharply [as shown in Fig.~\ref{Fig:1}(a)], which favors first-order phase transitions. Under the influence of AMMs, the asymmetry in the mass of $u-d$ quarks intensifies, thereby altering the phase boundaries. From the above discussion, especially in Fig.~\ref{Fig:2}, one finds that an appropriate value of the AMM induces IMC coupled with the magnetic field to suppress the critical temperature $T_{\textrm{c}}$, which is partially consistent with the results of LQCD ~\cite{Bali:2011qj, Ding:2020inp, Bali:2012zg, Ilgenfritz:2013ara, Bornyakov:2013eya, Bali:2014kia, tomiya2019phase}, but an unexpected first-order phase transition occurs.

\begin{figure}[H]
	\centering
	\includegraphics[width=0.85\textwidth]{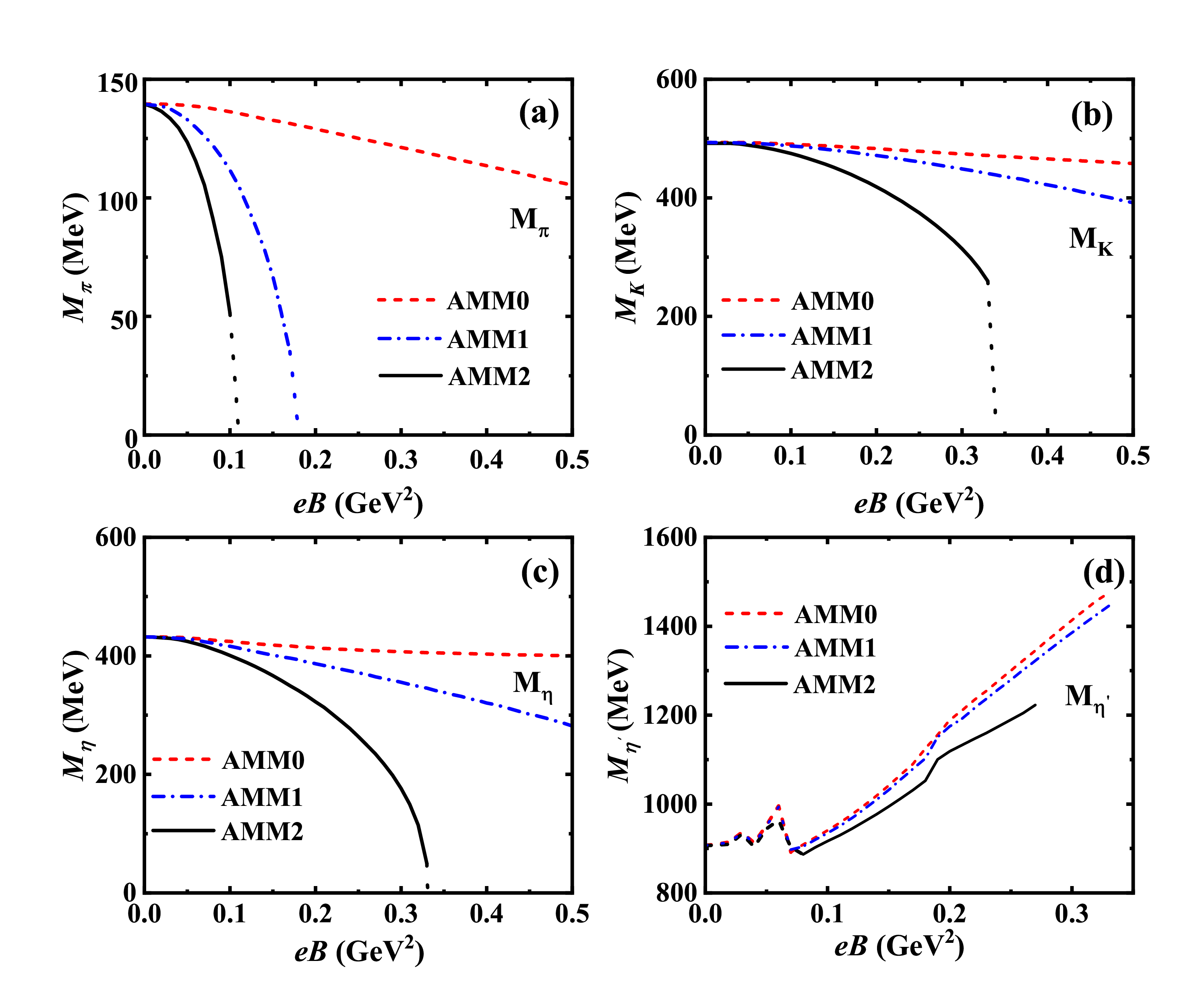}
	\caption{\label{Fig5} The magnetic field dependence of neutral pseudoscalar meson masses ($\pi$, $K$, $\eta$, $\eta'$) at zero temperature and chemical potential, comparing cases with and without the inclusion of AMMs. Panels (a), (b), (c), and (d) are for $\pi$,  $K$, $\eta$, and  $\eta'$, respectively.}
	\label{Fig:5}
\end{figure}

Figure~\ref{Fig:5} illustrates the magnetic field dependence of neutral pseudoscalar meson masses $(\pi,K,\eta,\eta')$ at zero temperature and chemical potential, comparing cases with and without the inclusion of quark AMMs. As shown in Fig.~\ref{Fig:5}(a), it is found that the $\pi$ meson mass decreases slightly with the increasing magnetic field ($eB$), reflecting mild suppression due to the MC effect without AMMs. But the $\pi$ mass drops sharply and collapses to zero at a critical magnetic field $eB_{\textrm{c}}\approx$ 0.17 GeV$^{2}$ with AMM1 and at $eB_{\textrm{c}}\approx$ 0.10 GeV$^{2}$ with AMM2, signaling a first-order phase transition of the pion mass. It is manifested that the $\pi$ meson, composed of light ($u/d$) quarks, is highly sensitive to AMMs. Although the variation of the u quark mass with the increasing magnetic field is not explicitly depicted in Fig. 1 (due to the analogous behavior of $u$ and $d$ quark masses governed by the gap equations), the presence of the magnetic field induces flavor symmetry breaking between $u$ and $d$ quarks, leading to an increasing mass splitting as the magnetic field increases. Notably, the inclusion of AMMs further exacerbates this flavor symmetry breaking. Combined with numerical results for pions, one can conclude that the behavior of the pion mass in vacuum strongly depends on the rate at which the magnetic field affects the flavor symmetry, in agreement with findings reported in Refs.~\cite{Xu:2020yag,Aguirre:2021ljk}. Additionally, the study of the effective coupling of the neutral pion to quark-antiquark pairs also plays some role in understanding the magnetic field dependence of pion properties. This aspect is not described in the present work, and also the detailed studies can be founded in Ref.~\cite{Aguirre:2021ljk}.

As shown in Fig.~\ref{Fig:5}(b), it is found that the $K$ meson mass decreases gradually but remains finite across the studied $eB$ range without considering the AMM and with considering AMM1. But the $K$ mass declines more steeply and vanishes at $eB_{\textrm{c}}\approx$ 0.33 GeV$^{2}$ with AMM2. It is manifested that the $s$ quark component of $K$ meson reduces its sensitivity to AMMs compared to $\pi$. According to recent LQCD studies on meson masses~\cite{Ding:2020hxw}, the observed reduction in neutral meson masses under increasing magnetic field may be linked to the IMC mechanism. Numerous LQCD simulations have reported that MC dominates at zero temperature. In this work, in scenarios excluding the AMM or incorporating moderate AMM1, the masses of $\pi$ and $K$ mesons exhibit a mild decrease due to the influence of the dominant MC effect. The inclusion of small AMM1 also manifests MC effects, with IMC being secondary, and thus suppresses the chiral condensate and consequently also the strength of chiral symmetry breaking(shown in Fig.~\ref{Fig:1}). As a result, the mass decrease is more obvious in the case with the AMM1 compared to the case without AMM, which seems to further support the LQCD reports that the decrease in neutral meson mass may be associated with IMC.

However, we note a more surprising case with a larger AMM (AMM2 set): even at zero temperature, AMM2 directly induces IMC, leading to a significantly accelerated decline in meson masses and even triggering a mass collapse. Prior to this, two NJL model studies also reported the collapse of pion mass, where Ref.~\cite{Xu:2020yag} employed a soft cutoff scheme and Ref.~\cite{Aguirre:2021ljk} utilized an improved-MFIR regularization scheme applicable to the AMM. Additionally, it is crucial to emphasize that within the studied magnetic field range, the cases without AMMs or with AMM1 are acceptable results and consistent with LQCD predictions for mass decline~\cite{Ding:2020hxw}. In contrast, the AMM2 case shows a surprising feature in mass, which make it less eligible. Moreover, together with the $\eta$ meson mentioned below, one finds that the larger the meson mass is, the less the meson mass is affected by the magnetic field, which has also been reported in lattice QCD simulations~\cite{Ding:2020hxw}.

The $\eta$ mass decreases slowly, maintaining a smooth trend with AMM0 and AMM1 as shown in Fig.~\ref{Fig:5}(c). But a rapid mass reduction occurs with AMM2, culminating in a sharp drop to zero at $eB_{\textrm{c}}\approx$ 0.33 GeV$^{2}$, similar to the $K$ meson. It is manifested that the $\eta$ meson mixed flavor structure (the $s$ quark component) makes its mass less sensitive to quark AMMs. AMM-induced isospin symmetry breaking amplifies $\eta$ coupling to the magnetic field, leading to a faster fall, similarly reported for pion and $K$, as mentioned above. The AMM affects the masses of meson by enhancing the splitting of quark masses, especially for states dominated by light flavors ($\pi$, $K$, $\eta$).

As shown in Fig.~\ref{Fig:5}(d), the $\eta'$ mass increases monotonically with $eB$ but becomes undefined above $eB_{c}\approx$ 0.33 GeV$^{2}$ due to divergences in polarization loops without AMMs. Even with the introduction of AMMs, the $\eta'$ mass still increases, but the magnitude is suppressed, and the solution is lost at a lower $eB_{c}\approx$ 0.27 GeV$^{2}$ for AMM2. The $\eta'$ meson, regarded as a resonance state in the NJL model even for $T$ = 0 due to the lack of confinement~\cite{Rehberg:1995kh}, exhibits strong sensitivity to the $\eta'$ decay width. The presence of the AMM makes this limitation worse by enhancing nonperturbative decay effects, thus restricting its physical solutions in strong fields. Additionally, the NJL model loses its validity at higher energies. At zero temperature, the mass of the $\eta'$ meson slightly exceeds the cutoff scale $\Lambda$ , or in other words, the mass of the $\eta'$ meson lies in a region where the NJL model is close to being invalid, resulting in insufficient predictive capability for the $\eta'$ meson compared to other neutral mesons. For these reasons, the $\eta'$ meson does not exhibit mass collapse, as it may fail to support the formation of a physical mass solution as magnetic fields increase.

To summarize briefly, the differences caused by the AMM are obvious, like the observed collapse of the pion mass for AMM1 and distinct behavior of the thermal susceptibility $\chi_{T}$. These differences may also be attributed to our adoption of a constant AMM. To the best of our knowledge, the latest study~\cite{Fraga:2024klm} indicates that the quark AMM decreases with the increase of the magnetic field and even apparently vanishes for extremely large magnetic fields. Incorporating the magnetic- field-dependent AMM $\kappa_f(eB)$ into the calculations, the pion mass might not vanish at a certain critical magnetic field or would only vanish at a higher magnetic field. This could potentially reconcile the differences observed in Fig.~\ref{Fig:1}(a),(b) and Fig.~\ref{Fig:5}(a). Moreover, this issue guides us toward deeper investigation in the future. One can expect that calculations of the quark AMM from lattice QCD will serve as an important input parameter. Similar to the fitting of the magnetic- field-dependent coupling constant $G(eB)$ within the NJL model, we aim to obtain a magnetic field-dependent AMM $\kappa_f(eB)$. In this way, implementing magnetic-field-dependent $\kappa_f(eB)$ instead of a constant $\kappa_f$ in the NJL model framework will allow us to explore this area rigorously and to see if any novel results emerge.
\begin{figure}[H]
	\centering
	\includegraphics[width=0.35\textwidth]{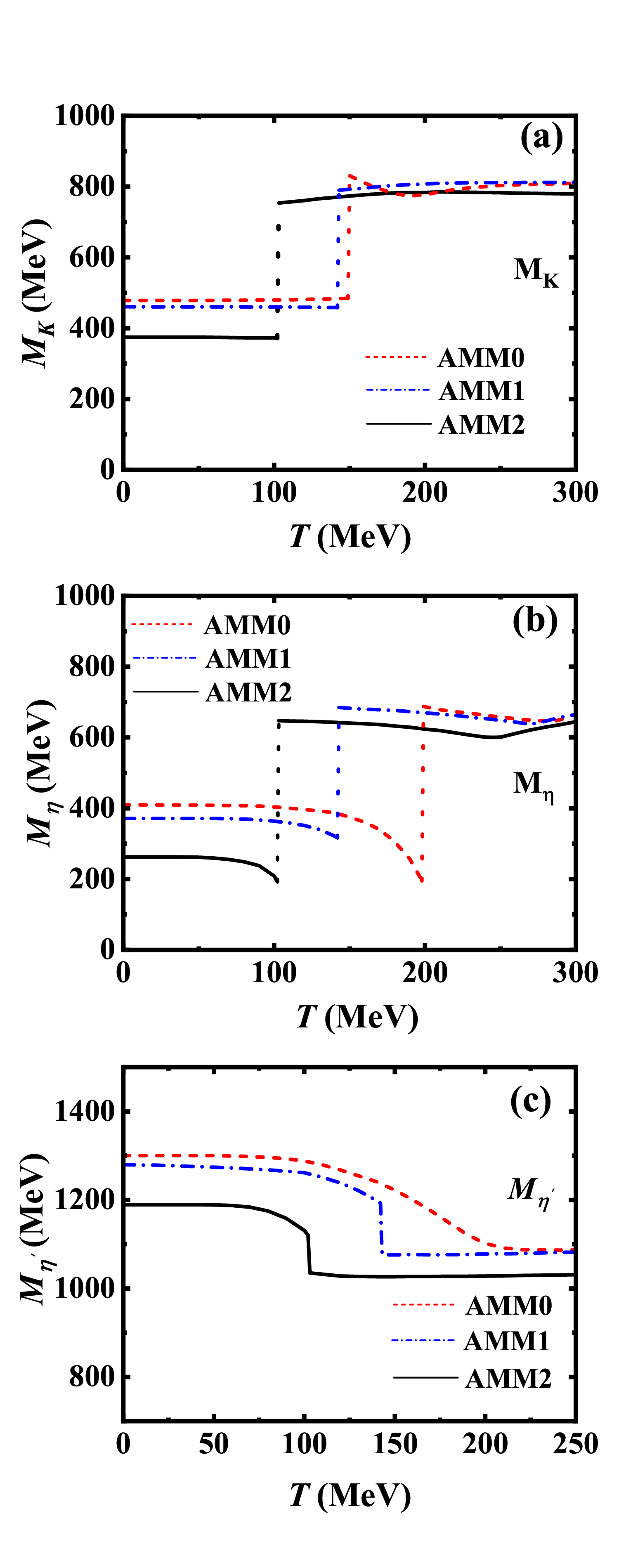}
	\caption{\label{Fig6} $K$, $\eta$, and $\eta'$ meson masses as a function of temperature without and with considering AMMs at zero chemical potential. Panels (a), (b), and (c) are for  $K$, $\eta$, and $\eta'$, respectively.}
	\label{Fig:6}
\end{figure}

Figure~\ref{Fig:6} depicts the temperature dependence of neutral pseudoscalar meson masses ($K$,$\eta$, and $\eta'$) under a constant magnetic field ($eB_{\textrm{c}}$ = 0.25 GeV$^{2}$), comparing scenarios with and without AMMs, at zero chemical potential. As shown in Fig.~\ref{Fig:6}(a), the $K$ meson mass increases slightly with temperature until reaching the Mott temperature ($T_{\textrm{mott}}\approx$ 149 MeV), where it undergoes a discontinuous mass jump without considering AMMs. This marks the transition from a bound state to a resonance state. Beyond $T_{mott}$ , the mass first decreases slightly, then rises again at $T\sim$ 190 MeV. Numerical results obtained with the smaller AMM1 closely resemble those of AMM0. But when considering the larger AMM2, the $K$ mass decreases slightly with temperature at low  $T$ ($<$ 102 MeV), contrasting the AMM-free case. A sharp mass jump occurs at the first-order phase transition temperature of light quarks ($u/d$), followed by a gradual increase until $T\sim$ 216 MeV, after which the mass declines for AMM2. The larger AMM significantly affects the $k$ meson mass by coupling light quark dynamics to the magnetic field, and the first-order quark mass drop directly triggers the $K$ mass jump.

As shown in Fig.~\ref{Fig:6}(b), it is found that the $\eta$ mass decreases slowly with temperature until $T_{\textrm{mott}}\approx$ 198 MeV, where a mass jump occurs without considering AMMs. Then, the mass continues to decline before rising again at $T\sim$ 284 MeV. But when considering AMMs, the $\eta$ mass jumps abruptly at the light quark first-order transition temperature, followed by a gradual decrease. For example, at $T\sim$ 247 MeV for AMM2, the mass begins to rise. At high temperature, the $\eta$ mass is systematically lower than in the AMM-free case. The $\eta$ meson mixed flavor composition (combining light and strange quarks) makes it sensitive to both $u/d$ quark phase transitions and AMM-enhanced magnetic effects. The AMM shifts the mass jump to a lower temperature, reflecting stronger chiral symmetry restoration.

As shown in Fig.~\ref{Fig:6}(c), it is found that the $\eta'$ mass decreases monotonically with temperature until $T\sim$ 235 MeV without considering AMMs, after which it begins to rise. But when considering AMMs, the $\eta'$ mass decreases initially, exhibits a sudden drop at the light quark first-order transition, then rises slightly at $T >$ 164 MeV for AMM1 and at $T >$ 151 MeV for AMM2. The overall mass reduction is more pronounced compared to the AMM-free case. Similarly as a resonance state in the NJL model, the $\eta'$ is highly sensitive to its resonance state decay widths. The AMM amplifies nonperturbative effects, destabilizing the $\eta'$ through enhanced decay widths.
\begin{figure}[H]
	\centering
	\includegraphics[width=0.35\textwidth]{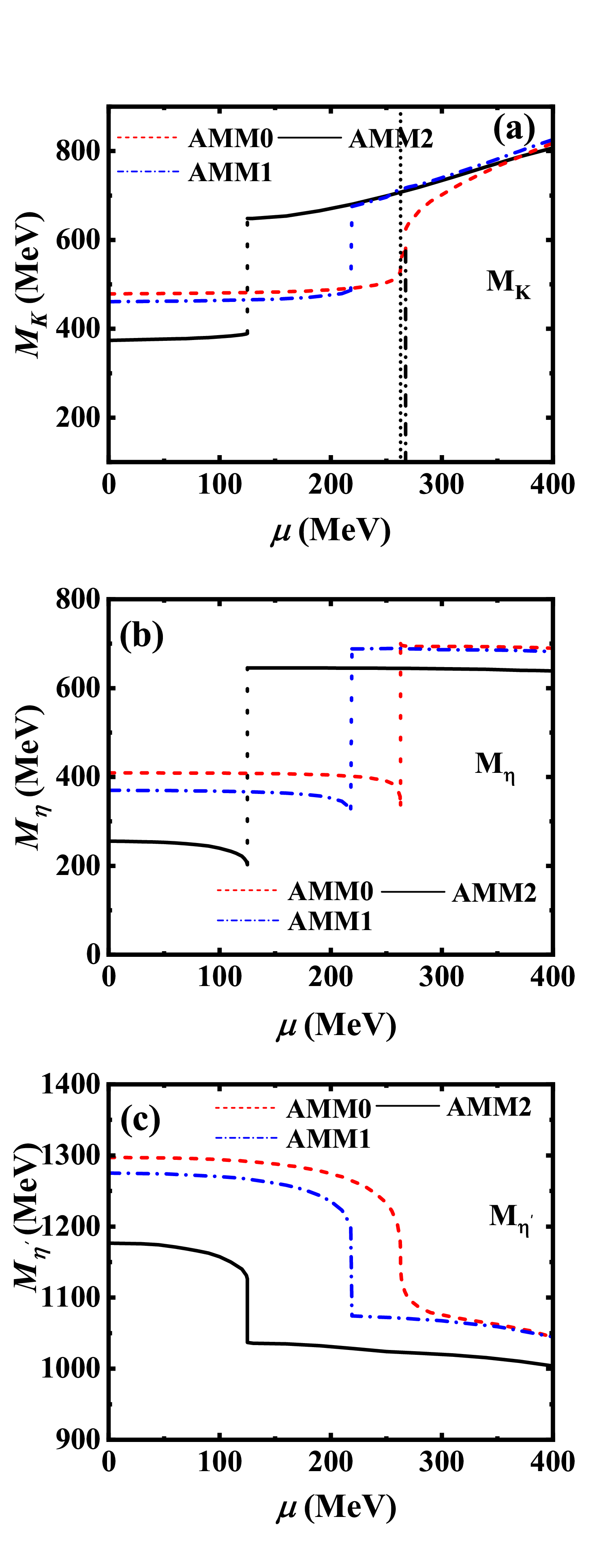}
	\caption{\label{Fig7} The dependence of the masses of neutral pseudoscalar mesons ($K$, $\eta$ and $\eta'$) on the quark chemical potential ($\mu$) under a fixed temperature ($T$ = 75 MeV) and magnetic field ($eB_{\textrm{c}}$ = 0.25 GeV$^{2}$). Panels (a), (b), and (c) are for $K$, $\eta$, and $\eta'$, respectively.}
	\label{Fig:7}
\end{figure}
Figure ~\ref{Fig:7} illustrates the dependence of the masses of neutral pseudoscalar mesons ($K$, $\eta$, and $\eta^{'}$) on the quark chemical potential ($\mu$) under a fixed temperature ($T$ = 75 MeV) and magnetic field ($eB_{\textrm{c}}$ = 0.25 GeV$^{2}$), comparing scenarios with and without the inclusion of AMMs. As shown in Fig.~\ref{Fig:7}(a), it is found that the $K$ meson mass increases continuously with $\mu$, exhibiting a steep rise near the CEP at $\mu_{\textrm{CEP}}\approx$ 262.8 MeV without considering AMMs. A Mott transition (discontinuous jump) occurs at $\mu > \mu_{\textrm{CEP}}$, marking the transition from a bound state to a resonance state. But when considering AMMs, the $K$ meson undergoes an abrupt Mott transition directly at $\mu_{\textrm{CEP}}(\textrm{AMM1})\approx$ 218.0 MeV for AMM1 and at $\mu_{\textrm{CEP}}(\textrm{AMM2})\approx$ 124.9 MeV for AMM2. Consistent with the finite temperature case, this shift highlights that IMC induced by the AMM at finite chemical potential can be seen as a catalyst to accelerate chiral symmetry restoration.

As shown in Fig.~\ref{Fig:7}(b), it is found that the $\eta$ meson mass decreases smoothly with $\mu$, interrupted by a mass jump at $\mu_{\textrm{CEP}}$ without considering AMMs. But when considering AMM1 and AMM2, the mass jump shifts to $\mu_{\textrm{CEP}}(\textrm{AMM1})$ and $\mu_{\textrm{CEP}}(\textrm{AMM2})$, followed by a continued decline. The mixed flavor composition of $\eta$ (combining light and strange quarks) makes it less sensitive to both $u/d$ quark phase transitions and AMM-enhanced magnetic effects. The AMM amplifies isospin symmetry breaking, resulting in lower mass compared to AMM0.

As shown in Fig.~\ref{Fig:7}(c), it is found that the $\eta'$ meson mass decreases monotonically with $\mu$, with the steepest decline aligning with $\mu_{\textrm{CEP}}$ without considering AMMs. But when considering AMM, a sudden mass drop occurs at $\mu_{\textrm{CEP}}(\textrm{AMM})$, followed by a gradual decline. Similarly as a resonance state, $\eta'$ is highly sensitive to its decay widths. The AMM amplifies nonperturbative effects, destabilizing $\eta'$ through enhanced decay widths.

The shift from smooth crossovers to first-order transitions under AMMs demonstrates how quark-level magnetic interactions reshape the QCD phase diagram. This provides insights into the competition between thermal, density, and magnetic effects in chiral symmetry restoration. By reproducing features like IMC predicted by numerical LQCD, the inclusion of AMMs improves the predictive power of effective models like NJL. It bridges gaps between theory and experiments, guiding future refinements for charged and vector meson studies.

\section{SUMMARY AND CONCLUSIONS}\label{sec:5}

In this study, we systematically investigated the effects of quark AMMs on the mass spectra of neutral pseudoscalar mesons ($\pi$, $K$, $\eta$, $\eta'$) under external magnetic fields, finite temperatures, and quark chemical potentials within the framework of the three-flavor NJL model. By incorporating AMMs at the quark level, we explored the interplay between magnetic catalysis, inverse magnetic catalysis, and chiral symmetry restoration, addressing discrepancies between conventional NJL predictions, i.e., not considering AMMs.

Key findings from our analysis include (1) quark mass dynamics: The inclusion of AMMs significantly altered the magnetic field dependence of constituent quark masses. For light quarks ($u/d$), a first-order phase transition emerged at critical magnetic fields ($eB_{\textrm{c}}\approx$ 0.33 GeV$^{2}$), where their dynamical masses collapsed to current quark masses with a larger AMM. In contrast, the strange quark exhibited nonmonotonic mass behavior but avoided abrupt transitions. These results highlight the critical role of AMMs in suppressing chiral condensates(or quark masses by the gap equation in the NJL model) and driving IMC. (2) Phase diagram modifications: the introduction of AMM reshaped the QCD phase structure. At zero and finite chemical potential ($\mu$), AMM suppressed the chiral phase transition temperature ($T_{\textrm{c}}$), aligning more closely with LQCD findings on IMC compared to the NJL model without considering AMMs (AMM0). Moreover, crossover transitions observed in the absence of AMMs were re replaced by first-order transitions with AMMs under strong magnetic fields. (3) Meson mass behavior: The $\pi$ mass collapsed abruptly at $eB_{\textrm{c}}\approx$ 0.17 GeV$^{2}$ with AMM1 and at $eB_{\textrm{c}}\approx$ 0.10 GeV$^{2}$ with AMM2. AMMs enhanced flavor splitting, leading to a significant impact on meson masses. As a resonance state, $\eta'$ showed suppressed mass growth under AMMs and lost stability at strong fields, reflecting limitations of the NJL framework in handling nonperturbative decay widths. (4) Temperature and chemical potential dependence: at finite temperature and chemical potential, meson masses underwent a Mott transition (discontinuous mass jump) at $T_{mott}$ (e.g., $K$ and $\eta$). The AMM-induced IMC shifted the Mott temperature toward lower temperatures. Regarding chemical potential dependence, the AMM-induced IMC also caused the mass jump of mesons to shift toward lower chemical potential.

In terms of experiments, chiral symmetry breaking gave rise to rich mesonic and baryonic spectra, while the $U_{A}(1)$ anomaly explained the nondegeneracy of the $\eta$ and $\eta'$ mesons. The mass changes of mesons may enhance or suppress their thermal production in relativistic heavy-ion collisions, such as the yields and ratios of mesons. The Mott transition may result in some fascinating phenomena in relativistic heavy-ion collisions where strong magnetic fields are generated.

Our results demonstrated that the AMM plays a pivotal role in reconciling the NJL model predictions with LQCD observations, particularly in reproducing IMC for transition temperature. However, challenges remain in fully capturing the $\eta'$ meson behavior, underscoring the need to incorporate decay widths and beyond-RPA corrections. Future work will extend this framework to charged and vector mesons, which are crucial for understanding electromagnetic probes in heavy-ion collisions and magnetar environments. Additionally, refining the interplay between AMMs and the $U_{A}(1)$ anomaly could further elucidate the chiral and axial restoration mechanisms in magnetized QCD matter.

\begin{acknowledgements}
	This work was supported by the National Natural Science Foundation of China (Grants No. 11875178, No. 11475068, and No. 11747115).
\end{acknowledgements}

\appendix
\section{THE CALCULATION OF POLARIZATION LOOPS}\label{appendix:Cal_HighEnergyLimit}
\setcounter{equation}{0}
\renewcommand\theequation{A\arabic{equation}}

In this appendix, we derive the main part of the polarization function containing the quark AMMs at finite temperature. We start with the following expression using kaon meson as an example:
\begin{equation}\label{eq:1}
	\begin{aligned}
		\int \frac{d^4 k}{(2\pi)^4} Tr \{ \gamma^5 S_d(q = p + k) \gamma^5 S_s(k) \}.
	\end{aligned}
\end{equation}
Substitute Eq. (21) into Eq. (A1) and by doing the trace operation on the gamma matrix, we get,
\begin{equation}\label{eq:2}
	\begin{aligned}
	&\int \frac{d^4k}{(2\pi)^4} \mathrm{Tr}\left\{\gamma^5 S_d (q = p + k) \gamma^5 S_s(k)\right\} \\
	&= \int \frac{d^4k}{(2\pi)^4} \, 2e^{-\frac{q_\perp^2}{|e_d B|} - \frac{k_\perp^2}{|e_s B|}} \sum_{n,m,s,l} (-1)^{n+m} \frac{1}{M_{nd} M_{ms}} \left[(-q_0 k_0 + q_z k_z) + (s M_{nd} - \kappa_d e_d B)(l M_{ms} - \kappa_s e_s B)\right] \\
	&\quad \times \Bigg\{ (M_{nd} + s M_d)(M_{ms} + l M_s) L_n \left( \frac{2q_\perp^2}{|e_d B|} \right) L_m \left( \frac{2k_\perp^2}{|e_s B|} \right) \\
	&\quad + (M_{nd} - s M_d)(M_{ms} - l M_s) L_{n-1} \left( \frac{2q_\perp^2}{|e_d B|} \right) L_{m-1} \left( \frac{2k_\perp^2}{|e_s B|} \right)  \\
	&\quad+ 8ls(q_x k_x + q_y k_y) L_{n-1}^1 \left( \frac{2q_\perp^2}{|e_d B|} \right) L_{m-1}^1 \left( \frac{2k_\perp^2}{|e_s B|} \right) \Bigg\} \times \\
	&\quad \left\{ \frac{1}{q_0^2 - E_{dns}^2 + i \epsilon} + 2 \pi i n_F (q_0) \delta (q_0^2 - E_{dns}^2) \right\} \left\{ \frac{1}{k_0^2 - E_{sml}^2 + i \epsilon} + 2 \pi i n_F (k_0) \delta (k_0^2 - E_{sml}^2) \right\}.
	\end{aligned}
\end{equation}
Now, considering $\vec{p}=0$, and using the orthogonality relation of the Laguerre function,
\begin{equation}\label{eq:3}
	\begin{aligned}
		\int_{0}^{\infty} dx e^{-x} x^{\alpha} L_{n}^{\alpha}(x) L_{m}^{\alpha}(x) = \frac{\Gamma(\alpha + n + 1)}{n!}, m = n, \operatorname{Re} \alpha > 0.
	\end{aligned}
\end{equation}
We arrive at
\begin{equation}\label{eq:4}
	\begin{aligned}
		&\int \frac{d^4 k}{(2\pi)^4} Tr \{ \gamma^5 S_d(q = p + k) \gamma^5 S_s(k) \} \\
		&= -\frac{N_c}{(2\pi)^2} \sum_{n,s,l} \int \frac{dk_0 dk_z}{2\pi} \frac{1}{M_{nd}} \frac{1}{M_{ns}} \beta \left[ (-q_0 k_0 + k_z k_z) + (sM_{nd} - \kappa_d e_d B)(lM_{ns} - \kappa_s e_s B) \right] \times \\
		&\left[ (M_{nd} + sM_d)(M_{ns} + lM_s) + (M_{nd} - sM_d)(M_{ns} - lM_s) + 4ls\beta n \right] \times \\
		&\left\{ \frac{1}{q_0^2 - {E_{dns}}^2 + i\varepsilon} + 2\pi i n_F(q_0) \delta(q_0^2 - {E_{dns}}^2) \right\} \left\{ \frac{1}{k_0^2 - {E_{snl}}^2 + i\varepsilon} + 2\pi i n_F(k_0) \delta(k_0^2 - {E_{snl}}^2) \right\}.
	\end{aligned}
\end{equation}
Performing the $dk_{0}$ integral of Eq.(A4), we get
\begin{equation}\label{eq:5}
	\begin{aligned}
		&\Pi^{ps}_{K^0}(p_0 = m_{K^0}) = \frac{\beta N_c}{2(2\pi)^2} \sum_{n,s,l} (1 + sl \frac{M_d M_s + 2\beta n}{M_{nd} M_{ns}}) \\&
		\left\{ \int dk_z \frac{1}{E_{dns}} \left[ 1 - \frac{1}{e^{(E_{dns} - \mu)/T} + 1} - \frac{1}{e^{(E_{dns} + \mu)/T} + 1} \right] + \int dk_z \frac{1}{E_{snl}} \left[ 1 - \frac{1}{e^{(E_{snl} - \mu)/T} + 1} - \frac{1}{e^{(E_{snl} + \mu)/T} + 1} \right] \right.\\&
		\left. + \left\{ [(M_{nd} - s\kappa_d e_d B) - sl(M_{ns} - l\kappa_s e_s B)]^2 - p_0^2 \right\} B(m_d, m_d) \right\},
	\end{aligned}
\end{equation}
where,
\begin{equation}\label{eq:6}
	\begin{aligned}
		&B(m_d, m_s) =\\& \int dk_z \left\{ \frac{1}{E_{dns}} \left[ \frac{1}{(E_{dns} + p_0)^2 - {E_{snl}}^2} \frac{1}{e^{-(E_{dns} + \mu)/T} + 1} - \frac{1}{(E_{dns} - p_0)^2 - {E_{snl}}^2} \frac{1}{e^{-(E_{dns} - \mu)/T} + 1} \right] \right. \\&
		+ \left. \frac{1}{E_{snl}} \left[ \frac{1}{(E_{snl} - p_0)^2 - {E_{dns}}^2} \frac{1}{e^{-(E_{snl} + \mu)/T} + 1} - \frac{1}{(E_{snl} + p_0)^2 - {E_{dns}}^2} \frac{1}{e^{(E_{snl} - \mu)/T} + 1} \right] \right\}.
	\end{aligned}
\end{equation}

\bibliography{Feng}
\end{document}